\documentclass{aa}
\usepackage{natbib,graphics}
\usepackage{times}
\bibpunct{(}{)}{;}{a}{}{,}

\newcommand{\footiraf}{\footnote{IRAF is distributed by the National
    Optical Astronomy Observatories, which is operated by the Association
    of Universities for Research in Astronomy, Inc. (AURA) under
    cooperative agreement with the National Science Foundation}}

\begin{document}

\sloppy
  
\title{Spectral Types of Planetary Host Star Candidates: \\ Two New Transiting
  Planets?}
\author{S. Dreizler\inst{1}, T. Rauch\inst{1,2}, 
  P. Hauschildt\inst{3}, S.L. Schuh\inst{1}, W. Kley\inst{4}, K. Werner\inst{1}}

\offprints{Dreizler, \\ {\tt dreizler@astro.uni-tuebingen.de}}

\institute{Institut f\"ur Astronomie und Astrophysik, Abt. Astronomie, Sand 1,
           Universit\"at T\"ubingen, 
           D-72076 T\"ubingen, Germany 
           \and
           Dr.\,Remeis-Sternwarte, Sternwartstra\ss e 7, D-96049 Bamberg, Germany
           \and
           Dept. of Physics \& Astronomy, The University of Georgia,
           Athens, GA 30602-2451, USA
            \and Institut f\"ur Astronomie und Astrophysik,
           Abt. Computational Physics, Auf der Morgenstelle 10, Universit\"at T\"ubingen, D-72076
           T\"ubingen, Germany 
}
\date{Received 7 June 2002 / Accepted 9 July 2002}
\authorrunning{Dreizler et al.}
\titlerunning{Spectral Types of Planetary Host Star Candidates:Two New
  Transiting Planets? } 
\abstract{Recently, 46 low-luminosity object transits were reported from
  the Optical Gravitational Lensing Experiment. Our follow-up spectroscopy
  of the 16 most promising candidates provides a spectral classification of the
  primary. Together with the radius ratio from the transit measurements, we
  derived the radii of the low-luminosity companions. This allows to
  examine the possible sub-stellar nature of these objects. Fourteen of them
  can be clearly identified as low-mass stars. Two objects,
  \object{OGLE-TR-03} and \object{OGLE-TR-10} have companions with radii of
  0.15\,R$_\odot$ which is very similar to the radius of the transiting planet
  HD\,209458\,B. The planetary nature of these two objects should therefore
  be confirmed by dynamical mass determinations.
 } \maketitle

\section{Introduction}
The detection of planets outside our solar system was a longstanding goal
of astronomy. After the first detections
(\citealt{latham:89,wolszczan:92,mayor:95}), an intensive search with
various methods began (see \citealt{schneider:02} for an overview). Out of
the currently 102 known planets, 100 have been detected with Doppler velocity
measurements of the planets host stars. All these planets were found around
solar like stars. The other two are planets around pulsars and were found
by periodic pulse modulation measurements.

The Doppler method is subject to several selection effects which are
problematic for a more general understanding of planet formation and
evolution. It is mainly applied to solar like stars (spectral type F---K)
because they provide sufficient lines to measure the radial velocity with
the required precision of the order of m/sec.
Radial velocity detections favor close-in and
massive planets.  Therefore, many Jovian planets are found within
Mercury-like orbits.  Regardless of the selection effects, the detection of
extra-solar planets has already had a large impact on the understanding and
evolution of planetary systems. Establishing a less biased sample would, however, be a
big step forward.

No planet has yet been found by photometric monitoring. The (currently)
unique planetary companion of HD\,209458 has an orbital inclination which
allows the measurement of the eclipse of the host star by the planet
\citep{charbonneau00,henry:00}. This planetary companion was, however, known before
from Doppler measurements \citep{mazeh:00}. Recently 46 transiting planet
candidates were 
announced by the OGLE (Optical Gravitational Lensing Experiment) consortium
\citep{udalski:02}. These candidates were extracted from a sample of about
5 million stars observed during a 32-day photometric monitoring. In a
sub-sample of 52\,000 stars with a photometric accuracy better than 1.5\%,
these 46 candidates exhibit light curves indicating the presence of a
transiting low-luminosity companion. From the analyses of the light curves,
the radii of the visible primaries and of the invisible secondaries were
derived. Up to now, no spectroscopic information of the primary is
available. The goal of this project is to provide this information
and to infer the nature of these low-luminosity companions.

We will describe the observations, data reduction and discuss
the determination of the spectral types of the primaries in Sect.\, 2. The
results are discussed  in Sect.\, 3.

\section{Observations, Data Reduction, and Spectral Types of the Primary Stars}
We selected 16 candidates from the list of \citet{udalski:02}, 13 of
  these have the smallest predicted companion radii.
The spectra were obtained as back-up program by one of
us (T.R.) at the 
SAAO 1.9\,m telescope using the Grating Spectrograph equipped with a
266$\times$1798 SITe chip. This follow-up will be continued to complete the
  whole list of \citet{udalski:02}.
The grating\,7 provides a spectral resolution of
5\,\AA, exposure times were set to 1800\,sec for all objects. Standard
  data 
  reduction of these long-slit spectra was performed using IRAF\footiraf
  and included bias subtraction, flat-field correction as well as
  wavelength and flux calibration. Comparing the targets from the
list of \citet{udalski:02}, we note that \object{OGLE-TR-08} is identical
to \object{OGLE-TR-29}.

\begin{figure}[th]
\vspace{6.7cm}
\includegraphics{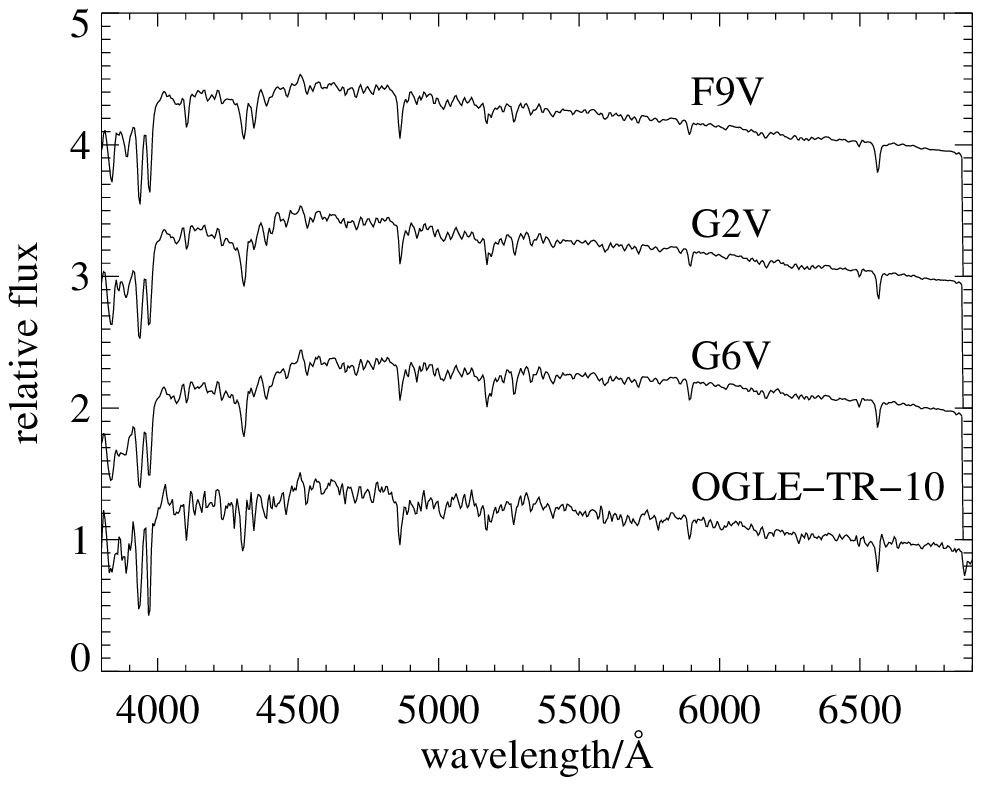}
\caption[]{OGLE-TR-10 (bottom) compared to three template spectra with the
  best matching one in the middle. Main differences are the strengths of
  the Balmer lines and of the G-Band.} 
\label{Ferror}
\end{figure}
\begin{figure*}[th]
\vspace{13.5cm}
\includegraphics{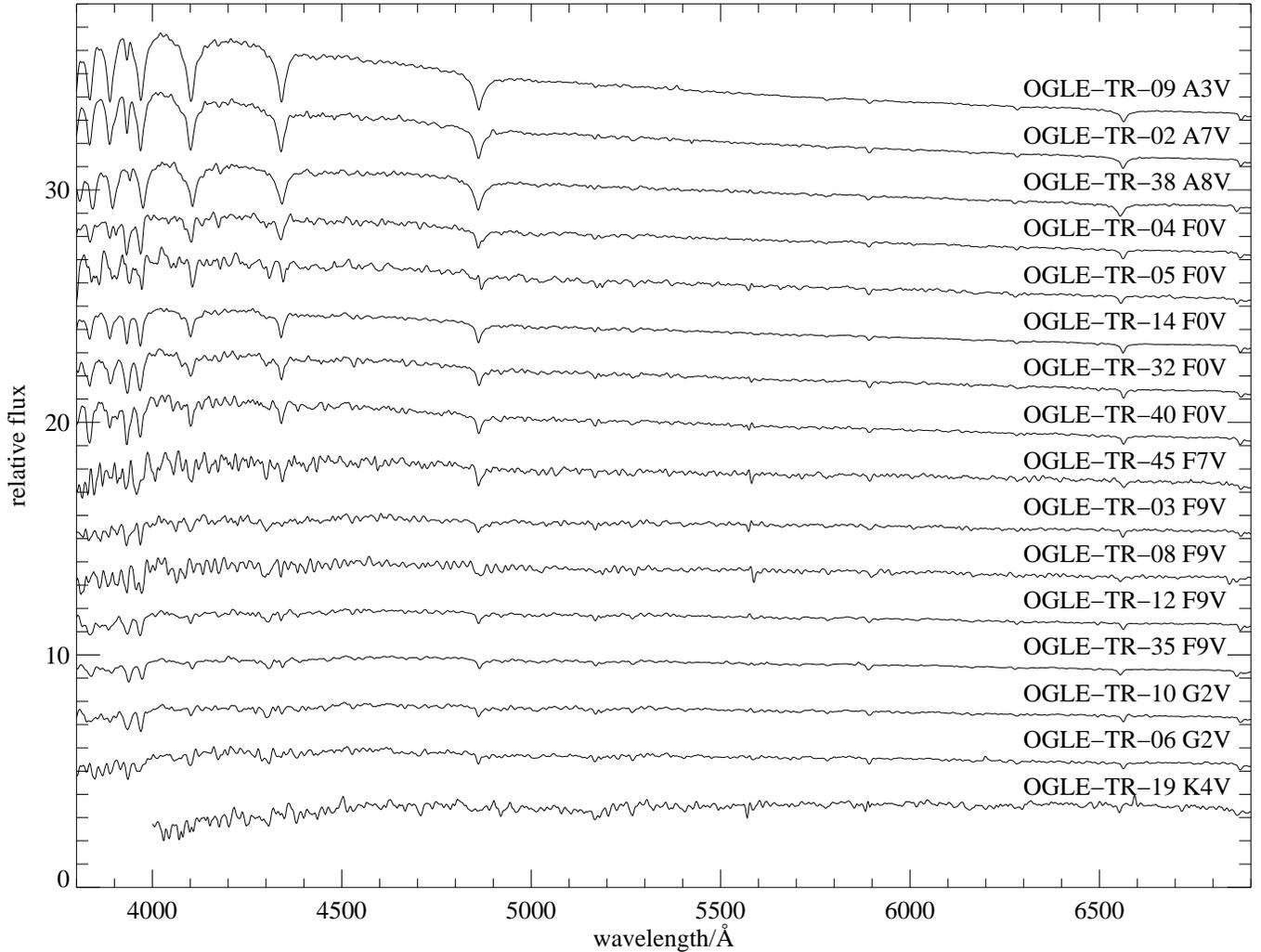}
\caption[]{Spectra of our target stars with our spectral
  classification. The spectral types cover stars from the maximum down to
  vanishing Balmer lines. Also visible is the maximum strength of the Ca H
  and K doublet as well as the increasing strength of the G-band. We do not
  find spectral signature of the low-luminosity companion.} 
\label{Fall}
\end{figure*}

The obtained spectra are compared to the spectral library of
\cite{silva:92} which provides templates in steps of about 0.3
  spectral classes. We use them without interpolation within the library. The quality
  of the match between observed and template spectrum is determined with a
  $\chi^2$ test. Re-binning the observed spectra and the templates to a
  common wavelength grid with 590 spectral bins, we obtain reduced $\chi^2$ close
  to unity for the best fits. Deviation in the $\chi^2$ from the best fit
  to the neighboring templates corresponds to deviations of more than
  3$\sigma$. The fitting therefore provides a classification better than half a spectral class.
In Fig.\,\ref{Ferror} we compare the most promising candidate
\object{OGLE-TR-10} with the best matching template and the next earlier and
later library spectrum. While e.g. hydrogen Balmer lines become too shallow in the
G6V template compared to the target star, they are too strong in the F9V
template. We restricted the classification to the luminosity class V since
the observed orbital periods indicate an orbital separation of the order of
ten solar radii and therefore prohibit the presence of a larger star. The
spectral classifications of all objects are displayed in Fig.\,\ref{Fall}.
The presence of the companion could not be detected from our data,
  neither from double lined spectra nor from the flux distribution.
%
\begin{table}[ht]
\caption{Light-curve variations \citep{udalski:02}, spectral types (SP), and
  derived quantities, i.e. primary radius, companion radius ratio, companion radius and mass in solar
  units, as well as mass ratio. The companion mass is derived assuming it
  is a low-mass star (thick lines Fig.\,\ref{Fmrr}). Typical error estimates for 
  derived companion radius and mass in the bottom part. See text for details.} 
\begin{tabular}{cccccccc}
\hline
$\#$ & $\!$ mmag$\!$ & $\!$SP$\!$ & $\!$R$_{\mathrm s}$/R$_{\odot}\!$ & $\!$R$_{\mathrm c}$/R$_{\mathrm s}\!$ &  $\!$R$_{\mathrm c}$/R$_{\odot}\!$ & $\!$M$_{\mathrm c}$/M$_{\odot}\!$ & $\!$M$_{\mathrm c}$/M$_{\mathrm s}\!$\\
\hline
 2   & 19        & A7 & 1.62          & 0.13            & 0.21                & 0.20  & 0.11\\
 3   & 19        & F9 & 1.14          & 0.13            & 0.15                & 0.13  & 0.12\\
 4   & 65        & F0 & 1.50          & 0.24            & 0.36                & 0.37  & 0.23\\
 5   & 43        & F0 & 1.50          & 0.20            & 0.30                & 0.30  & 0.19\\
 6   & 53        & G2 & 1.00          & 0.22            & 0.22                & 0.22  & 0.22\\
 8   & 48        & F9 & 1.14          & 0.21            & 0.24                & 0.23  & 0.21\\
 9   & 48        & A3 & 1.98          & 0.21            & 0.41                & 0.43  & 0.18\\
10   & 22        & G2 & 1.00          & 0.14            & 0.15                & 0.13  & 0.13\\
12   & 38        & F9 & 1.14          & 0.19            & 0.21                & 0.20  & 0.18\\
14   & 34        & F0 & 1.50          & 0.18            & 0.26                & 0.26  & 0.16\\
19   & 65        & K4 & 0.75          & 0.24            & 0.18                & 0.17  & 0.24\\
32   & 34        & F0 & 1.50          & 0.18            & 0.26                & 0.26  & 0.16\\
35   & 30        & F9 & 1.14          & 0.17            & 0.19                & 0.18  & 0.16\\
38   & 48        & A8 & 1.58          & 0.21            & 0.33                & 0.34  & 0.19\\
40   & 26        & F0 & 1.50          & 0.15            & 0.23                & 0.22  & 0.14\\
45   & 62        & F7 & 1.22          & 0.24            & 0.29                & 0.29  & 0.23\\
\hline
\hline
10   & 22        & F9 & 1.14          & 0.14            & 0.16                & 0.15  & 0.13\\
10   & 22        & G2 & 1.00          & 0.14            & 0.15                & 0.13  & 0.13\\
10   & 22        & G6 & 0.91          & 0.14            & 0.13                & 0.11  & 0.12\\
\hline
\label{Tspec}
\label{Terror}
\end{tabular}
\end{table}

We then used the derived spectral classes to estimate the stellar radii of
the primary stars (Tab.\,\ref{Tspec}) using the tabulated values from
\citet{allen}. The photometric monitoring of \citet{udalski:02} provides
the brightness variation during eclipses. Assuming a negligible radiation
from the secondary and a central passage in front of the primary this
brightness variation is directly proportional to the radius
ratio. Multiplied with the primary radius it yields the radius of the
secondary. Finally, we used the evolutionary models for low mass stars
(thick lines Fig.\,\ref{Fmrr}) to obtain the mass of the secondary assuming
it to be a 
low-mass star. The more sophisticated approach towards radius ratios,
i.e. to model the eclipse light curves with the derived primary radii as
constraint seems to be unnecessary with the current data set, because the
error for the companion radius is  dominated by the uncertainty of spectral
classification of the primaries and of the tabulated radii of the spectral
type standards.
Table\,\ref{Terror} additionally provides an estimate of the uncertainties in the
companion radius introduced by our spectral
classification. This error is small enough to obtain a quite clear picture
of the nature of the secondary star. 



\section{Discussion}
\begin{figure*}[th]
\vspace{10.0cm}
\includegraphics{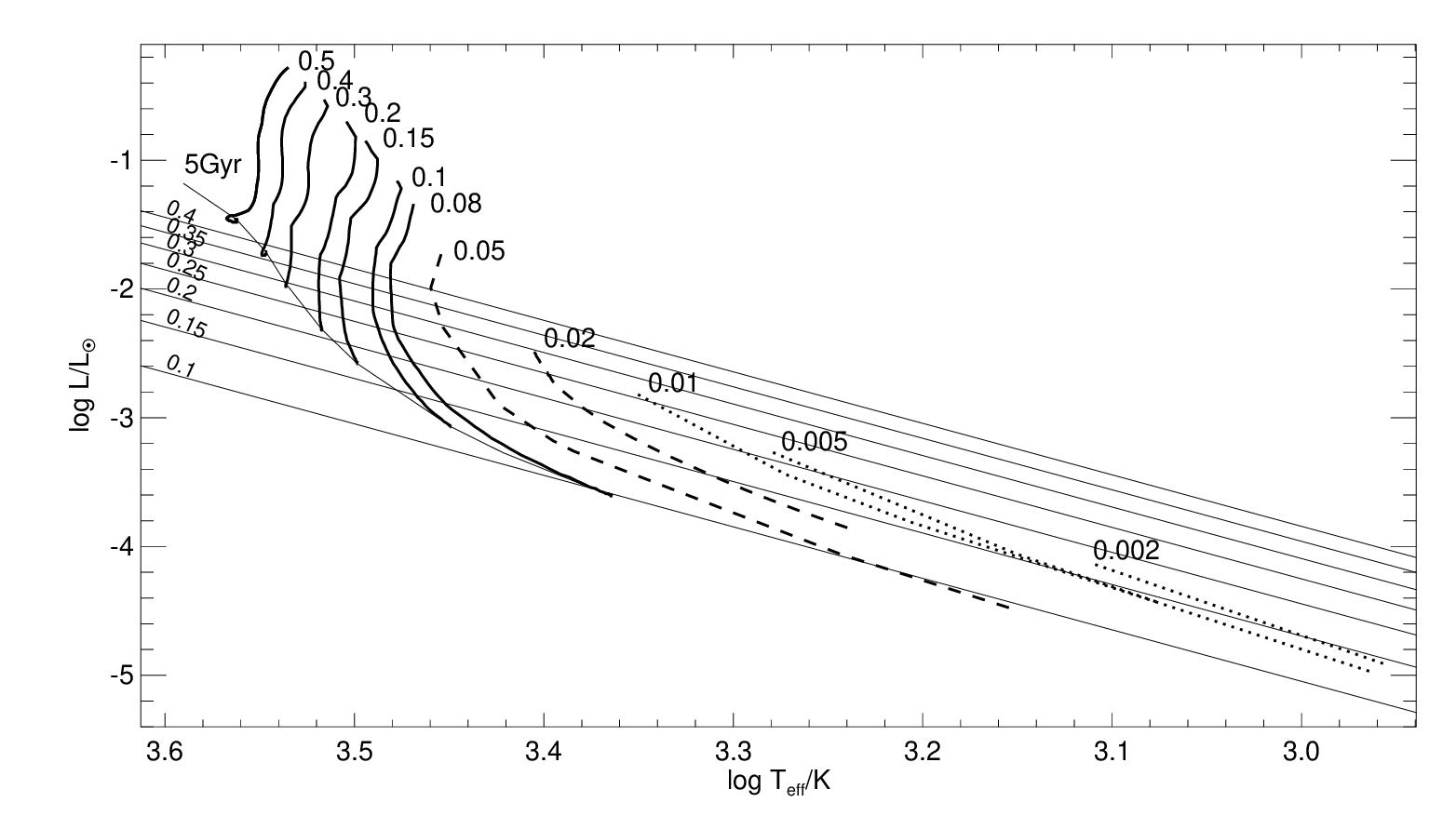}
\includegraphics{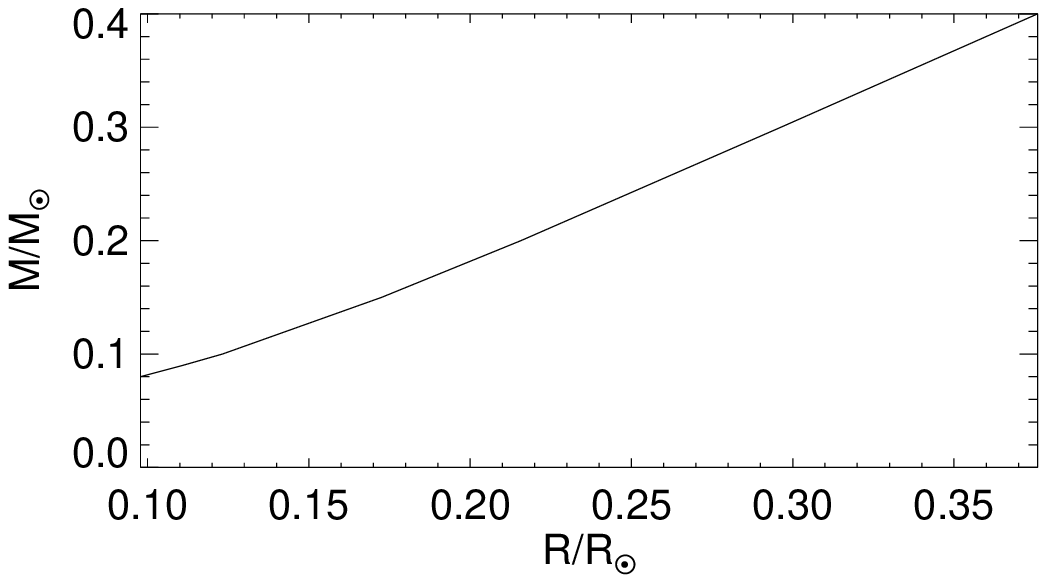}
\caption[]{Companion radii compared to evolutionary tracks of
   \citet{baraffe:98},
\citet{chabrier:00}, and \citet{baraffe:02} in the HRD. Thick lines:
  stellar models, dashed lines: brown-dwarf 
  models, dotted lines: gas-giant models. Note that the sub-stellar models
  are for {\em isolated} objects. Masses and radii are given in
  solar units. The inset figure shows the mass-radius relation for low-mass
  stars at an age of 5\,Gyr.}
\label{Fhrd}
\label{Fmrr}
\end{figure*}

The range of secondary radii is displayed in the Hertzsprung-Russell-Diagram
(Fig.\,\ref{Fhrd}) 
together with evolutionary tracks of \citet{baraffe:98},
\citet{chabrier:00}, and \citet{baraffe:02}.
The thick
lines indicate the (pre-)main sequence evolution of low-mass stars. The
position at an age of 5\,Gyr is indicated in the figure. We also display
the evolution of {\em isolated} contracting brown dwarfs (dashed) and gas
giants 
(dotted). The tracks of the sub-stellar models end at an age of 1\,Gyr for
0.05\,M$_\odot$ and 5\,Myr for 0.002\,M$_\odot$, respectively, and
therefore represent very young objects.

For the following discussion we assume that the OGLE-transits are
  undisturbed from blends of very nearby stars on the sky and that the
  transits are no grazing-incident eclipses. Even though these
  possibilities can not be completely ruled out, the former scenario seems
  unlikely because we do not detect an additional spectral contribution,
  the latter one because the photometry indicates flat-bottomed light
  curves.

All low-mass companions are found to have radii consistent with low-mass
stars of about M0V or later \citep{allen}. For all except two objects our relatively
large radii do not allow an interpretation as sub-stellar objects. This
list of low-mass star companions includes the best planetary companion
candidate, \object{OGLE-TR-40}, from \citet{udalski:02}, who derived a
companion radius of 0.1\,R$_\odot$. Modeling the eclipse light curve, they
derived a primary radius of 0.73\,R$_\odot$, which can be clearly
excluded from our spectroscopic determination. These systems are, however,
also interesting. As indicated in Tab.\,\ref{Tspec}, the mass ratio for
these binary stars is quite extreme. he formation of a close binary out of a
common proto-stellar disk favors typically a mass ratio of about unity
(e.g. \citealt{bate:97}). These 
low-mass objects in eclipsing binaries can also be used to calibrate the
mass-radius relation of these stars, providing constraints for evolutionary
models. This seems to be required since discrepancies are reported by
\citet{torres:02}.

For two objects, \object{OGLE-TR-03} and \object{OGLE-TR-10}, the derived
radius of 0.15\,R$_\odot$ does allow an interpretation as sub-stellar
objects. The latter was also among the two top candidates of
\citet{udalski:02}. In this case our spectroscopic determination fits
reasonably well with the light curve fit. In the case of
\object{OGLE-TR-03}, our radius is smaller than the one derived by
\citet{udalski:02}. 

Fig.\,\ref{Fhrd} shows that sub-stellar objects can be as
large as 0.15\,R$_\odot$, but only during a very early phase of their
evolution, i.e. 0.1\,Gyr for a 0.05\,M$_\odot$ brown dwarf and during
5\,Myr for a 0.002\,M$_\odot$ gas giant. It should be noted
that these tracks are calculated for {\em isolated} sub-stellar objects. The
separation of a few solar radii (derived from the orbital period and the
assumption that the companion mass is negligible) does indicate a strong
influence of the secondary. Theoretical models for sub-stellar companions
taking the irradiation of the primary into account are currently worked
on (e.g. \citealt{burrows:00}) and show that the large radii result
from the high residual entropy remaining from the early proximity of a
luminous companion. For the presently only known
transiting gas giant planet, HD\,209458B, this effect is indeed
observed. The derived radius is about 0.14\,R$_\odot$, despite the age of
probably several Gyrs.  
The
same is possible for \object{OGLE-TR-03} and \object{OGLE-TR-10}. While
\object{OGLE-TR-10} would be nearly a twin of the HD\,209458 system
regarding orbital period, spectral type of the primary, and companion
radius, \object{OGLE-TR-03} would be even more extreme. The orbital period
is only 1.18 days resulting in a separation of only 5.4\,R$_\odot$. In
combination with the earlier spectral type, the irradiation is even more
drastic. 

In summary, the spectroscopic follow-up of the most promising planetary
transit candidates did not result in a clear identification of a new
sub-stellar object, moreover most of the candidates could be identified as
low-mass stars. Two objects did, however, pass this spectroscopic test
and therefore continue to qualify as planetary candidates. The
ultimate determination of 
their nature does require a detailed study of radial velocity variations
with very high precision. Dynamical mass determination of the secondaries
with less demanding  
instrumental requirement will provide more insight in the mass-radius
relation at the lower end of the main sequence.

%

\begin{acknowledgements}
We use observations made at the South African Astronomical
Observatory (SAAO).
T.R. acknowledges a travel grant from the DFG
(RA 733/11-1).
This research was supported by the DLR under grant 50\,OR\,0201
(T\"ubingen).
\end{acknowledgements}

\end{document}